\documentclass[twocolumn,showpacs,aps,prd,superscriptaddress,letterpaper]{revtex4}

\usepackage{dcolumn}
\usepackage{epsfig}

\RequirePackage{xspace}

\input babarsym

\providecommand{\btodspi}{\mbox{$B^0\to D_{s}^+\pi^-$}}
\providecommand{\btodsosk}{\mbox{$B^0\to D_{s}^{(*)-} K^+$}}
\providecommand{\btodsspi}{\mbox{$B^0\to D_{s}^{*+}\pi^-$}}
\providecommand{\btodsk}{\mbox{$B^0\to D_{s}^{-}K^+$}}
\providecommand{\btodssk}{\mbox{$B^0\to D_{s}^{*-}K^+$}}
\providecommand{\btodsospi}{\mbox{$B^0\to D_{s}^{(*)+}\pi^-$}}
\providecommand{\btodsrho}{\mbox{$B^0\to D_{s}^+\rho^-$}}
\providecommand{\btodsosrho}{\mbox{$B^0\to D_{s}^{(*)+}\rho^-$}}
\providecommand{\btodssrho}{\mbox{$B^0\to D_{s}^{*+}\rho^-$}}
\providecommand{\btodskstar}{\mbox{$B^0\to D_{s}^-K^{*+}$}}
\providecommand{\btodskokstar}{\mbox{$B^0\to D_{s}^-K^{(*)+}$}}
\providecommand{\btodsoskstar}{\mbox{$B^0\to D_{s}^{(*)-}K^{*+}$}}
\providecommand{\btodsskstar}{\mbox{$B^0\to D_{s}^{*-}K^{*+}$}}
\providecommand{\btodsskokstar}{\mbox{$B^0\to D_{s}^{*-}K^{(*)+}$}}
\providecommand{\btodsoskokstar}{\mbox{$B^0\to D_{s}^{(*)-}K^{(*)+}$}}

\providecommand{\btodospi}{\ensuremath{B^0\to D^{(*)-}\pi^+}}

\providecommand{\btodsskstar}{\mbox{$B^0\to D_{s}^{*-} K^{*+}$}}

\providecommand{\btodskstar}{\mbox{$B^0\to D_{s}^{-} K^{*+}$}}

\providecommand{\dssdsgam}{\mbox{$D_{s}^{*+}\to D_{s}^+\gamma$}}

\providecommand{\dsphipi}{\mbox{$D^+_{s}\to \phi\pi^+$}}

\providecommand{\dsksk}{\mbox{$D^+_{s}\to \KS K^+$}}
\providecommand{\dskstark}{\mbox{$D^+_{s}\to \Kstarzb K^+$}}

\providecommand{\ds}{\ensuremath{D_{s}^+}}

\providecommand{\De}{\ensuremath{\Delta E}}

\providecommand{\lumi}{{\ensuremath{381\times10^6}}}

\providecommand{\stwobg}{\ensuremath{\sin(2\beta+\gamma)}\xspace}
\providecommand{\mDs}{\ensuremath{m(D_s)}}

\newcommand{\eqref}[1]{Eq.~(\ref{#1})}   % for referencing eqs by name

\newcommand{\BABARPubYear}    {08}
\newcommand{\BABARPubNumber}  {009}
\newcommand{\SLACPubNumber} {13190}

\long\def\inst#1{\par\nobreak\kern 4pt\nobreak
    {\it #1}\par\vskip 10pt plus 3pt minus 3pt}

\begin{document}

\begin{flushleft}
\babar-PUB-\BABARPubYear/\BABARPubNumber \\
SLAC-PUB-\SLACPubNumber \\
%hep-ex/\LANLNumber \\
%March 2008 \\
\end{flushleft}

% Title of the paper
\title{
Measurement of the Branching Fractions of the Rare Decays \\
\btodsospi, \btodsosrho, and \btodsoskokstar
}

% Input author list file
%% author list as of 05-Feb-2008 (544 authors)
%
\author{B.~Aubert}
\author{M.~Bona}
\author{Y.~Karyotakis}
\author{J.~P.~Lees}
\author{V.~Poireau}
\author{E.~Prencipe}
\author{X.~Prudent}
\author{V.~Tisserand}
\affiliation{Laboratoire de Physique des Particules, IN2P3/CNRS et Universit\'e de Savoie, F-74941 Annecy-Le-Vieux, France }
\author{J.~Garra~Tico}
\author{E.~Grauges}
\affiliation{Universitat de Barcelona, Facultat de Fisica, Departament ECM, E-08028 Barcelona, Spain }
% ...
\collaboration{The \babar\ Collaboration}
\noaffiliation

\date{March 29, 2008}

% Abstract
\begin{abstract}
We report the measurement of the branching fractions of the rare decays
\btodsospi, \btodsosrho, and \btodsoskokstar\ in a sample  of \lumi\
\FourS\ decays into \BB\  pairs collected with the \babar\
detector at the PEP-II asymmetric-energy $e^{+}e^{-}$ storage ring. 
We present evidence for the decay \btodskstar\ and the vector-vector
decays \btodssrho\ and  
\btodsskstar, as well as the first measurement of the vector meson polarization
in these decays. We also determine the ratios of the CKM-suppressed to
CKM-favored amplitudes $r(D^{(*)}\pi)$ and $r(D^{(*)}\rho)$ in
decays $\B^0\to D^{(*)\pm}\pi^{\mp}$ and $\B^0\to
D^{(*)\pm}\rho^{\mp}$, and comment on the prospects for measuring the
\CP\ observable $\sin(2\beta+\gamma)$. 
\end{abstract}

% PACS, the Physics and Astronomy Classification Scheme.

\maketitle

% The body of the paper starts here
\section{Introduction}

The Cabibbo-Kobayashi-Maskawa (CKM) quark flavor-mixing
matrix~\cite{CKM} provides an elegant explanation of the origin of
\CP\ violation within the Standard Model. 
\CP\ violation manifests itself as a non-zero area of the unitarity triangle~\cite{Jarlskog}.
While it is sufficient to measure one of the angles to demonstrate the
existence of \CP\ violation,
the unitarity triangle needs to be over-constrained by experimental
measurements 
in order to demonstrate that the CKM mechanism is the correct 
explanation of this phenomenon. Precision measurements of the sides
and angles of the unitarity triangle are the focus of the physics
program at the 
$B$ Factories. While several theoretically clean measurements
of the angle $\beta$ exist~\cite{sin2b}, constraining the other two
angles $\alpha$ and $\gamma$ is significantly more challenging. 
A theoretically clean measurement of $\sin(2\beta+\gamma)$ can 
be obtained from the study
of the time evolution for $\Bz\to D^{(*)-} \pi^+$~\cite{chconj}
and $\Bz\to D^{(*)-} \rho^+$
decays, which are available in large samples at the \B\ factories, and for the corresponding
CKM-suppressed modes $\Bz{\to} D^{(*)+}\pi^-$ and $\Bz{\to}
D^{(*)+}\rho^-$~\cite{sin2bg}. 
Measurements of
\CP\ asymmetries in decays $\Bz{\to} D^{(*)\mp}\pi^\pm$ and 
$\Bz{\to} D^{\mp}\rho^\pm$ decays have recently been
published~\cite{ref:s2bgDPi,ref:s2bgDRho}.

The interpretation of \CP\ asymmetries in
$\Bz{\to} D^{(*)\mp}\pi^\pm$ decays as a measurement of \stwobg\ 
requires knowledge of the ratios of the decay amplitudes,
\begin{equation}
r(D^{(*)}\pi)=\left|\frac{A(\Bz{\to} D^{(*)+}\pi^-)}
                           {A(\Bz{\to}D^{(*)-}\pi^+)}\right|\ . 
\label{eq:rDpi}
\end{equation}
However, direct measurements of the doubly Cabibbo suppressed
branching fractions 
$\BR(\Bz{\to} D^{(*)+}\pi^-)$ and $\BR(\Bz\to D^{(*)+}\rho^-)$ are not 
possible with the currently available data samples due to the presence
of the copious background from $\Bzb{\to} D^{(*)+}\pi^-, D^{(*)+}\rho^-$.
On the other hand, assuming SU(3) flavor symmetry, 
$r({D^{(*)}\pi})$ can be related 
to the branching fraction of the decay \btodsospi~\cite{sin2bg}:
\begin{equation}
r(D^{(*)}\pi) = 
  \tan\theta_c\,
  \frac{f_{D^{(*)}}}{f_{D^{(*)}_s}}\sqrt{\frac{\BR(\btodsospi)}{\BR(\btodospi)}}
  \ ,
\label{eq:rDPi}
\end{equation}
where $\theta_c$ is the Cabibbo angle, and $f_{D^{(*)}}/f_{D^{(*)}_s}$
is the ratio of $D^{(*)}$ and $D^{(*)}_s$ meson decay
constants~\cite{fdsdTheory,fdsdRef,fdsdExp}. Other 
SU(3)-breaking effects are believed to affect $r({D^{(*)}\pi})$ by
(10-15)\%~\cite{ref:Baak}.

\begin{figure}[h]
\begin{center}
\epsfig{file=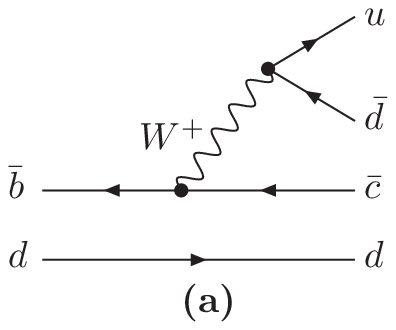,width=1.5in} 
\epsfig{file=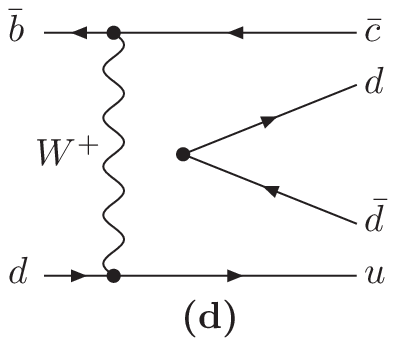,width=1.5in} \\ 
\epsfig{file=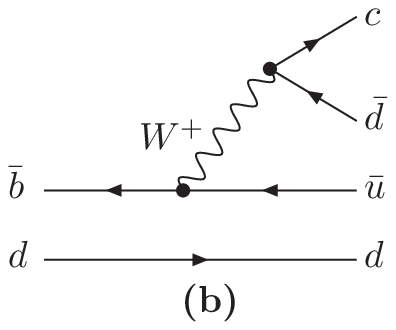,width=1.5in} 
\epsfig{file=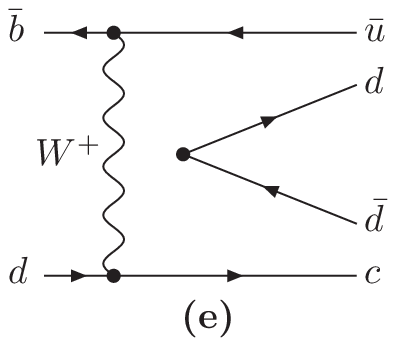,width=1.5in} \\
\epsfig{file=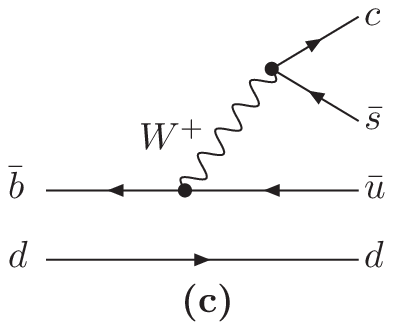,width=1.5in}
\epsfig{file=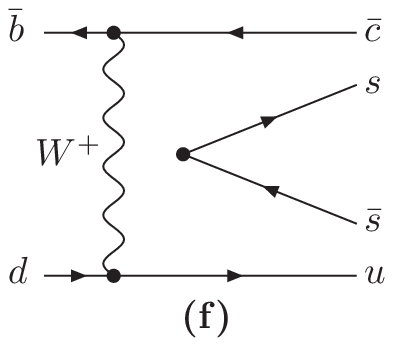,width=1.5in}
\end{center}
\vspace{-0.5cm}
\caption{Dominant Feynman diagrams for (a) CKM-favored decays
$\Bz\to D^{(*)-}\pi^+(\rho^+)$, (b) doubly CKM-suppressed decays
$\Bz\to D^{(*)+}\pi^-(\rho^-)$, and (c) the SU(3) flavor symmetry
  related decays $\Bz\to D_s^{(*)+}\pi^-(\rho^-)$;
(d) the color-suppressed $W$-exchange contributions
  to $\Bz\to D^{(*)-}\pi^+(\rho^+)$, (e) $\Bz\to D^{(*)+}\pi^-(\rho^-)$, and
(f) decay \btodsoskokstar.}    
\label{fig:diag} 
\end{figure}

The dominant Feynman diagrams for the decays 
$\Bz\to D^{(*)-}\pi^+(\rho^+)$, 
$\Bz\to D^{(*)+}\pi^-(\rho^-)$, 
$\Bz\to D_s^{(*)+}\pi^-(\rho^-)$,
and \btodsoskokstar\ are shown in Fig.~\ref{fig:diag}. 
Since  \btodsospi\ has four
different quark flavors in the final state, a single amplitude
contributes to the decay (Fig.~\ref{fig:diag}c). 
On the other hand, two diagrams contribute to 
$\Bz\to D^{(*)-}\pi^+$ and $\Bz\to D^{(*)+}\pi^-$:  
tree amplitudes (Fig.~\ref{fig:diag}a,b) and color-suppressed direct
$W$-exchange amplitudes (Fig.~\ref{fig:diag}d,e). The latter are
assumed to be negligibly small in Eq.~(\ref{eq:rDPi}). The decays \btodsosk\
(Fig.~\ref{fig:diag}f) probe the size of the $W$-exchange
amplitudes relative to the dominant processes $\Bz\to
D^{(*)-}\pi^+$. 

The rate of \btodsoskokstar\
decays could be enhanced by final state rescattering~\cite{Wexch}, in
addition to the $W$-exchange amplitude. Such long-distance effects
could also affect the vector meson polarization in \btodsskstar\
decays. 
The angular distribution in vector-vector decays 
$\Bz\to D_s^{*} V$ ($V=\rho,\, K^{*}$) is given by 
%%%%%%%%%%%%%%%%%%%%%%%
\begin{eqnarray}
\frac{d^2\Gamma}{d\cos\theta_{D_s^*}\,d\cos\theta_V} &\propto& 
\left[ (1 - f_L) (1+\cos^2\theta_{D_s^*})\sin^2\theta_V \right.
\nonumber\\
&+& \left. 4 f_L \sin^2\theta_{D_s^*}\cos^2\theta_V\right],
\label{eq:helicityshort}
\end{eqnarray}
%%%%%%%%%%%%%%%%%%%%%%%
\noindent 
where $\theta_{D_s^*}$ and $\theta_V$ are
the helicity angles of \Dss\ and the vector meson $V$,
respectively, 
$f_L=|A_0|^2/(\Sigma|A_\lambda|^2)$ is the
longitudinal polarization fraction, and
$A_{\lambda=-1,0,+1}$ are the helicity amplitudes. These distributions
are integrated over the angle between the decay planes of \Dss\ and
$V$. 

For amplitudes dominated by the short-range (electroweak)
currents, $f_L$ is 
predicted to be near unity~\cite{ref:VV}, with corrections of order 
$\mathcal{O}(m_V^2/m_B^2)$, where $m_V$ is the mass of the vector
meson produced by the weak current, and $m_B$ is the mass of the $B$
meson.  
Thus, the measurement of $f_L$ can constrain the size of the
long-distance contributions in \btodsskstar\ decays~\cite{Wexch}. 

The branching fractions $\BR(\btodsospi)$ and $\BR(\btodsosk)$ have been
measured previously 
by the \babar\ Collaboration~\cite{priorBaBar}. 
In this paper we present the first evidence for the decays
\btodssrho\ and \btodsoskstar, and a limit on the rate of \btodsrho. 
We also update the
measurements of the branching fractions $\BR(\btodsospi)$ and
$\BR(\btodsosk)$ with improved precision, using  a 65\%
larger dataset. 

\section{Data Sample and the Detector}

We use a sample of \lumi\ \FourS\ decays into \BB\
 pairs 
 collected with the \babar\ detector
at the \pep2\ asymmetric-energy \epem\ collider~\cite{pep}.
A detailed description of the \babar\ detector is available
 elsewhere~\cite{detector}. The components of the detector crucial to
 this analysis are  summarized below. 

Charged particle tracking is provided by a five-layer silicon
vertex tracker (SVT) and a 40-layer drift chamber (DCH). 
For charged-particle identification, ionization energy loss ($dE/dx$) in
 the DCH and SVT, and Cherenkov radiation detected in a ring-imaging
 device (DIRC) are used. 
Photons and neutral pions are identified and measured using
an electromagnetic calorimeter (EMC), which comprises 6580 thallium-doped CsI
crystals. These systems are mounted inside a 1.5-Tesla solenoidal
superconducting magnet. 
We use the GEANT4~\cite{geant4} software to simulate interactions of particles
traversing the \babar\ detector, taking into account the varying
detector conditions and beam backgrounds. 

\section{Event Selection and Analysis}
\label{sec:selection}

The selection of events of interest proceeds in two steps. First, we
preselect events with at least three reconstructed charged-particle tracks
and a total measured energy greater than $4.5$ GeV, as determined
using all charged particles and neutral particles with energy above 30
MeV.
In order to reject $e^{+}e^{-}\to q\bar{q} (q=u,d,s,c)$ continuum
background, the ratio of the second to zeroth order Fox-Wolfram
moments~\cite{fox} must be less than $0.5$.

Candidates for \Ds\ mesons 
are reconstructed in the $\Ds\to\phi \pi^+$, $\KS K^+$ and
$\Kstarzb\Kp$ final states, with $\phi{\to} K^+K^-$, $\KS {\to} \pip \pim$, and
$\Kstarzb{\to} K^-\pi^+$.  
The $\KS$ candidates are reconstructed from two
oppositely-charged tracks, and their momentum is required to make an
angle $|\theta_\mathrm{flight}|<11^{\circ}$ 
with the line connecting their
vertex and the $\epem$ interaction point. All other tracks are  
required  to originate from the $\epem$ interaction region, loosely defined by
$|d_0|<1.5$~cm and $|z_0|<10$~cm, where $d_0$ and $z_0$ are the
distances of closest approach to the primary $\epem$ vertex in the
directions perpendicular and parallel to the beams, respectively.
In order to reject background from $\Dp{\to}\KS\pip$ or $\Kstarzb\pip$,
the $\Kp$ candidate in the reconstruction of $\Ds{\to}\KS\Kp$ or
$\Kstarzb\Kp$ is 
required to satisfy positive kaon identification criteria, which have an
efficiency of 85\%  and a 5\% pion misidentification probability. The same 
selection is used to identify kaon daughters of the
$B^0$ and $K^{*+}$ mesons in decays \btodsoskokstar. The selection is
based on the ratios of likelihoods for kaon, pion, and proton
identification in the SVT, DCH, and DIRC. The detector likelihoods are
calibrated over a wide range of momenta using particles identified
kinematically in clean decay chains, such as $D^{*+}\to D^0\pi^+$,
$D^0\to K^-\pi^+$, and $\Lambda\to p\pi^-$. 
In all other
cases, kaons are 
not positively identified, but instead candidates passing
a likelihood-based pion
selection are rejected. 
The selection efficiency of this ``pion veto'' is 95\% for the 
kaons and 20\% for the pions. 
Pion daughters of $B^0$ and $\rho^-$ mesons in the decays \btodsospi\
and \btodsosrho\ are required to be positively identified. 
Decay products of $\phi$, $\Kstarzb$, $\Ds$, and $\Bz$ candidates are 
constrained to originate from a single vertex. 

%
% rho and K*
%
%
% rho selection
%
We reconstruct $\rho^+\to\pi^+\pi^0$ candidates by combining a
well-identified charged pion with a $\pi^0\to\gamma\gamma$
candidate. The $K^{*+}$ candidates are reconstructed through the
decays $K^{*+}\to K^+\pi^0$ and $K^{*+}\to K^{0}_S\pi^+$. 
The neutral pion candidates are reconstructed from a pair of photons
each with a minimum energy of $30$~MeV. The invariant mass of
the photon pair is
required to be within $\pm 25$~\mevcc of the nominal
value~\cite{PDG2006}. The selected candidates are
constrained to the nominal $\pi^0$ mass before forming the
$\rho^+$ or $K^{*+}$ candidates. 
We require that
the invariant mass of the two 
pions forming the $\rho^-$ candidate be within $\pm 320$~\mevcc\ of the nominal
value~\cite{PDG2006}, and the invariant mass of the $K^+\pi^0$ and
$K^{0}_S\pi^+$ pairs be within $\pm 75$~\mevcc\ of the nominal
$K^{*+}$ mass~\cite{PDG2006}. $K^{0}_S\pi^+$ pairs are
constrained to a common geometric vertex.

%%RF HERE GO THE CRITERIA FOR THE D_S* 
We reconstruct  \Dss\ candidates in the mode  $\Dss{\to}\Ds\gamma$ by
combining  \Ds\ and photon candidates.
Photon candidates are required to be consistent with an electromagnetic
shower in the EMC, and to have an energy greater than $100$~\mev in the
laboratory frame. 
When forming a \Dss, the \Ds\ candidate is required to have an invariant mass
within 10~\mevcc\ of the nominal value~\cite{PDG2006}. For \btodssrho\
and \btodsskstar\ modes, we apply a ``$\pi^0$ veto'' by rejecting
photons that in combination with 
any other photon in the event form an  invariant mass that
falls within $125< m_{\gamma\gamma}<145$~\mevcc. 

%
% Likelihood selection
%
The efficiency of the initial preselection discussed above varies
between $14\%$ (\btodssrho, \dskstark) and $48\%$ (\btodspi,
\dsphipi). After the preselection, we identify signal $B$ decay
candidates using a likelihood ratio  
$R_L =
\mathcal{L}_\mathrm{sig}/(\mathcal{L}_\mathrm{sig}+\mathcal{L}_\mathrm{bkg})$,
where
$\mathcal{L}_\mathrm{sig}=\prod_{j}\mathcal{P}_{\mathrm{sig}}(x_k)$
is the multivariate likelihood for 
the signal hypothesis and
$\mathcal{L}_\mathrm{bkg}=\prod_{i}\mathcal{P}_{\mathrm{bkg}}(x_k)$
is the likelihood for 
the background hypothesis. Here $x_k$ represents one of the discriminating
variables described below, which are computed for each event. 
The likelihoods for the signal and background hypotheses are
computed as a product of the probability density functions (PDFs)
$\mathcal{P}_{\mathrm{sig}}(x_k)$ and
$\mathcal{P}_{\mathrm{bkg}}(x_k)$, respectively, for 
the following selection variables: invariant masses of the $\phi$,
$\Kstarzb$, $\rho^+$, $K^{*+}$, and $\KS$ candidates; $\chi^2$
confidence level of the vertex 
fit for the $\Bz$ and $\Ds$ mesons; the helicity angles of the $\phi$,
$\Kstarzb$, $\rho^+$, $K^{*+}$, and $\Dss$ meson decays; the
mass difference 
$\Delta m(\Dss) = m(\Dss)-m(\Ds)$; the polar angle $\theta_B$ of the
$B$ candidate 
momentum vector with respect to the beam axis in the
$\epem$ center-of-mass (c.m.) frame; the angle $\theta_{T}$ 
between the thrust axis of  the  $B$ candidate and the thrust
axis of all other particles in the event in the c.m. frame; 
the event topology variable $\mathcal{F}$, and the kinematic variable
\De, described below.  
Correlations among these variables are small. 

The helicity angle
$\theta_H$ is defined as the angle between one of the decay products  of
a vector meson and the flight direction of its parent particle in the meson's
rest frame. Polarization of the vector mesons
in the signal decays causes the cosines of their helicity angles to be
distributed as 
$\cos^2\theta_H$ ($\phi$, $\Kstarzb$, $\rho^+$, and $K^{*+}$) or
$1-\cos^2\theta_H$ (\Dss), while the random background combinations tend to 
produce a more uniform distribution in $\cos\theta_H$, with a peak in
the forward direction (which corresponds to a low-energy $\pi^0$) for
$\rho^+$ and $K^{*+}$ candidates. 
We do not include the helicity angles for \Dss, $\rho^+$, and
$K^{*+}$ mesons in the likelihood ratio $R_L$ for the vector-vector
\btodssrho\ and \btodsskstar\ modes, since the polarizations of the
vector mesons in these decays are not known {\em a priori}. Instead,
the helicity angles are used in the multi-dimensional likelihood fit
to determine the polarizations, as discussed below. 

The variables $\cos\theta_B$, $\cos\theta_T$, and $\mathcal{F}$
discriminate between spherically-symmetric \BB\ events and jet-like
continuum background using event topology. 
The polar angle $\theta_B$ is distributed as $\sin^2\theta_B$
for real $B$ decays, while being nearly flat in $\cos\theta_B$ for
the continuum.
\BB\ pairs form a nearly uniform $|\cos\theta_T|$
distribution, while the $|\cos\theta_T|$ distribution for 
continuum events peaks at 1.
A linear (Fisher)
discriminant $\mathcal{F}$ is
derived from the values of sphericity and thrust for the event, 
and the two Legendre moments $L_0$ and
$L_2$ of the energy flow 
around the $B$-candidate thrust axis~\cite{ref:legendre}.

\begin{figure}
\begin{center}
\epsfig{file=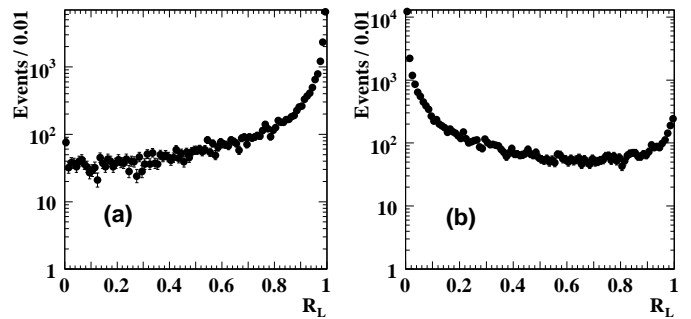,width=3.5in}
\end{center}
\vspace{-0.5cm}
\caption{Distribution of the likelihood ratio $R_L$ for the mode
  \btodspi, \dsphipi. Shown are (a) the simulated signal events, and
  (b) the sum of the simulated background samples from the $B^0$
  and $B^+$ decays, and $\epem\to q\bar{q}$ events.
}
\label{fig:likeCut} 
\end{figure}

The ratio $R_L$ has a maximum at $R_L=1$ for
signal events, and at $R_L=0$ for background originating
from continuum events. It also discriminates
well against $B$ decays without a real $\Ds$ meson in the final
state. 
The Monte Carlo (MC) simulated distributions of the $R_L$ variable for
signal and 
background events, in \btodspi\ decays, are shown in
Fig.~\ref{fig:likeCut}. 

Finally, two other variables \mes\ and $\De$ take advantage of the unique
kinematic properties 
of the $\epem\to\Upsilon(4S)\to \BB$ decays. The beam energy spread
is significantly smaller than the energy resolution of the
reconstructed $B$ mesons, and at the same time larger than the
momentum resolution. The momentum of the signal
candidates is included in the beam-energy-substituted mass 
$\mes = \sqrt{ (s/2 +
  \mathbf{p}_{i}\cdot\mathbf{p}_{B})^{2}/E_{i}^{2}- 
  \mathbf{p}^{2}_{B}}$, 
where $\sqrt{s}$ is the total c.m.
energy,  $(E_{i},\mathbf{p}_{i})$ is the four-momentum of the initial
\epem\ system, and $\mathbf{p}_{B}$ is the \Bz\ candidate momentum,
both measured in the laboratory frame. The second variable is 
$\De = E^{*}_{B} - \sqrt{s}/2$, where $E^{*}_{B}$ is the \Bz\ candidate
energy in the c.m. frame. 
For signal events, the \mes\ distribution is nearly Gaussian and centered
at the $B$ meson mass with a resolution of about ($2.5$-$2.8$)~\mevcc,
and the \De\ distribution has a 
maximum near zero with a resolution of (17-25)~\mev. We include \De\
in the definition of the likelihood ratio $R_L$; \mes\ is
used as a discriminating variable in the maximum likelihood fit
described below.

We parameterize the signal and background PDFs using large samples
of simulated events.  
We select \btodsospi\ and \btodsosk\ candidates that satisfy
$R_L>0.85$, and accept \btodsosrho\ and \btodsoskstar\ candidates with
$R_L>0.96$. We measure the relative efficiencies $\varepsilon_{R_L}$ of
the $R_L$ selections in copious 
data samples of decays $\Bz\to D^-\pi^+,\, D^-\rho^+$ ($D^-\to
K^+\pi^-\pi^-,\, \KS\pi^-$) 
and $B^+\to\Dbar^{*0}\pi^+,\,\Dbar^{*0}\rho^+$ ($\Dbar^{*0}\to
\Dbar^0\gamma,\, D^0\to K^-\pi^+$) 
in which the kinematics is similar to those of our signal events, and find
that they are consistent with Monte Carlo estimates of
$\varepsilon_{R_L}\approx 70\% (40\%)$ for \btodsospi\ and \btodsosk\
(\btodsosrho\ and \btodsoskstar) modes. The fraction of continuum
background events passing the selection varies between $2\%$ and $15\%$,
depending on the mode. 

%
% CHECK THESE NUMBERS !!! (Checked 02/14/08)
%
Less than 30\% of the selected events in the \btodsspi, \btodssk,
\btodsosrho, and \btodsoskstar\
channels ($<2\%$ in \btodspi\ and \btodsk)
contain two or more 
candidates that satisfy the criteria listed above. In
such events we select a single $B^0$ candidate based on
an event $\chi^2$ formed from \De, \mDs\ and (where appropriate) $\Delta
m(\Dss)$, $m_\rho$, $m_{K*}$, $m_{\pi^0}$ and $m_{Ks}$, and their 
uncertainties. 

%
%%% Likelihood fit
%
\section{Extraction of Signal Yields}

After the $R_L$ requirement is applied, we define the region of
interest using the
beam-energy-substituted mass \mes\ and the mass of the \Ds\ candidate \mDs. 
We require 
$5.2<\mes<5.3$~\gevcc and 
$|\mDs-\mDs_\mathrm{PDG}|<50$~\mevcc\ for \btodspi, \btodsrho, and
\btodskokstar\ modes, where 
$\mDs_\mathrm{PDG}$ is the world average $D_s$ mass~\cite{PDG2006}. 
The invariant
mass \mDs\ has a resolution of
(5-6)~\mevcc, depending on the $\Ds$ decay mode. 
The selection is significantly broader than the region populated by
the signal events, and allows us to constrain backgrounds in the signal
region. 
For \btodsspi, \btodssrho, and \btodsskokstar, we require
$|\mDs-\mDs_\mathrm{PDG}|<10$~\mevcc.

Five classes of background events contribute to the fit region. 
First is the {\em combinatorial background\/}, in which a true or fake
$D_s^{(*)}$ candidate is combined with a randomly-selected light
meson. 
Second,  $B$ meson decays such as  $\Bzb{\to}D^{(*)+}\pim$ or 
$\Bzb{\to}D^{(*)+}\rho^-$ with
$\Dp{\to}\KS\pip$ or $\Kstarzb\pip$ can 
constitute a background for  the \btodsospi\ and \btodsosrho\ modes if
 the pion in the $D$ decay is misidentified as a kaon ({\it{reflection
background}}). The reflection background has nearly the same \mes\ distribution
as the signal but different distributions in \De\ and \mDs.
The corresponding backgrounds for the
\btodskokstar\ mode ($\Bz{\to}\Dm K^{(*)+}$) are negligible. 
Third, rare {\em charmless\/} $B$  decays into the same final state,
such as $\Bz{\to} \Kbar^{(*)0}\Kp h$ (where $h=\pi,\,\rho,\,K$, or
$K^*$), 
have the same \mes\ and \De\ distributions as the $\Bz\to D_s h$
signal, but are nearly flat in \mDs. The charmless background is 
significant in \btodspi, \btodsrho, and \btodskokstar\ decays, but is
effectively rejected by the $\Delta m(\Dss)$
variable for the modes with \Dss.

Finally, two classes of background events have nearly the same distribution
as the signal events in both \mDs\ and \mes. For \btodsoskstar\
modes we take into account the potential contributions from 
the {\em non-resonant\/} decays $\Bz\to D_s^{(*)-}K^0\pi^+$ (which have
recently been observed~\cite{ref:Vitaly}), and 
color-suppressed $\Bz\to D_s^{(*)-}K^+\pi^0$ (unobserved so
far). Analogous non-resonant modes $B^0\to D_s^{(*)+}\pi^-\pi^0$
require the additional popping of a color-matched $q\bar{q}$ pair. They are
expected to be small compared to \btodsosrho~\cite{ref:Vitaly} and
are ignored.  Finally, 
{\em crossfeed background\/} from misidentification of 
$\Bbar^0\to D_s^{(*)-}\pi^+$ events
as \btodsosk\ signal, and vice versa, needs to be taken into account.

%%% yield extraction
For each mode of interest, we perform an unbinned extended 
maximum-likelihood (ML) fit to separate the signal events from the
backgrounds and extract the signal branching fractions. 
For \btodspi, \btodsrho, \btodsk, and \btodskstar, we perform a
two-dimensional fit to the \mes\ and \mDs\ distributions. 
For \btodsspi\ and \btodssk\ decays, we fit the 
one-dimensional \mes\ distribution. In vector-vector modes
\btodssrho and \btodsskstar, we constrain both the branching fractions of
the signal modes and the polarization of the vector mesons by
performing a three-dimensional fit to the distribution of \mes, and
the two helicity angles of the $\Dss$ and $\rho^-$ ($K^{*+}$) mesons. 

For each $B$ decay, we simultaneously fit distributions in the three 
\Ds\ decay modes,  
constraining the signal branching fractions to a common value. 
The likelihood function contains the contributions 
of the signal and the five background components discussed above. 
The function to be maximized is
%%%%%%%%%%%%%%%%%%%%%%%
\begin{equation}
{\cal L} = \exp\left(-\sum_{k,m}^{} n_{km}\right)\,
\prod_{i=1}^{N_{\rm cand}}
\left(\sum_{j}~n_{jm}\,
{\cal P}_{jm}(\vec{\zeta}_{i})\right)
\label{eq:likel}
\end{equation}
%%%%%%%%%%%%%%%%%%%%%%%
where $n_{jm}$ is 
the number of events for each event type $j$
(signal and all background modes) in each $D_s$ decay mode $m$, 
and ${\cal P}_{jm}(\vec{\zeta}_{i})$ is the 
probability density function of the variables
$\vec{\zeta}_{i}=(\mes, \mDs, \cos\theta_{D_s^*}, \cos\theta_V)$
for the $i$th event. The likelihood product is computed over all
candidates $N_\mathrm{cand}$ in the region of interest. We
parameterize the event yields as 
\begin{equation}
n_{jm} = N_{\BB} \mathcal{B}_j\mathcal{B}_m^{Ds}\varepsilon_m\, ,
\label{eq:yieldDef}
\end{equation}
where $m$ stands for $\dsphipi$, $\dskstark$, or $\dsksk$, 
$N_{\BB}=\lumi$, $\mathcal{B}_j$ is the $B$ decay branching fraction,
$\mathcal{B}_m^{Ds}$ is the branching fraction of the 
$m$-th $\ds$ mode, and $\varepsilon_m$ is the reconstruction efficiency. 

The branching fractions of the channels contributing to the reflection
background  are fixed in the
fit to the current world average values~\cite{PDG2006} and the
branching fractions of the crossfeed backgrounds are determined by
iterating the fits over each $B$ decay mode. 
The branching fractions of the non-resonant backgrounds are fixed to
the values recently measured by \babar~\cite{ref:Vitaly}. In the case of
$\Bz\to D_s^{(*)-}K^+\pi^0$, which can contribute to \btodsoskstar\
($K^{*+}\to K^+\pi^0$), we estimate the branching fraction by
%%%%%%%%%%%%%%%%%%%%%
\begin{eqnarray}
\BR(\Bz\to D_s^{(*)-}K^+\pi^0) \approx & \nonumber\\
\BR(B^+\to D_s^{(*)-}K^+\pi^+) &
\frac{\BR(B^0\to\Dzb\pi^0)}{\BR(B^+\to\Dzb\pi^+)}\, .
\label{eq:DsKpi0}
\end{eqnarray}
%%%%%%%%%%%%%%%%%%%%%
This scaling assumes that the dominant mechanism for producing
both $D_s^{(*)-}K^+\pi^0$ and $D_s^{(*)-}K^+\pi^+$ final states is a
sub-threshold production of a charmed $D^{**0}$ meson, which
subsequently decays into $D_s^{(*)-}K^+$, as indicated by the
invariant mass spectrum of $D_s^{(*)-}K^+$~\cite{ref:Vitaly}. Since 
the decay $\Bz\to D^{**0}\pi^0$ is color-suppressed compared to 
$B^+\to D^{**0}\pi^+$, we estimate the color
suppression factor from the $B^0\to\Dzb\pi^0$ decays. Direct
production of the color-suppressed $D_s^{(*)-}K^+\pi^0$ final state
(without the intermediate $D^{**0}$) results in a smaller branching
fraction estimate. We assign a 100\% systematic uncertainty to 
$\BR(\Bz\to D_s^{(*)-}K^+\pi^0)$. 

The expected yields of the dominant $B$-decay
backgrounds are listed in Table~\ref{tab:bkg}. 

The PDFs and efficiencies for the signal, reflection, and crossfeed
backgrounds are determined independently for each \Ds\ decay mode
using Monte Carlo samples. 
The signal contribution is modeled as a Gaussian (\btodspi\ and \btodsk) 
or a ``Crystal Ball'' function~\cite{ref:CB} in \mes\ and a double
Gaussian in \mDs. 
The combinatorial background is 
described in \mes\ by a threshold function~\cite{argus}, 
$dN/dx\propto x\sqrt{1-2x^{2}/s}\exp\left[-\xi\left(1-2x^{2}/s\right)\right]$,
characterized by the shape parameter $\xi$. 
This shape
parameter, common to all \Ds\ modes, is allowed to vary in the fit. 
In \mDs, the combinatorial background is well described by a
combination of a first-order polynomial (fake \Ds\ candidates) and a
Gaussian with (5-6)~\mevcc\ resolution (true \Ds\
candidates). The charmless background is parameterized by the signal
Gaussian shape in \mes\ and a first order polynomial in \mDs. 

%
% HELICITY DISTRIBUTIONS
%
Ideally, the distribution of the helicity angles in the vector-vector
decays is given by Eq.~(\ref{eq:helicityshort}). 
The helicity angle $\theta_{D_s^*}$ is defined as the angle between
the direction of 
the photon in $D_s^*\to D_s\gamma$ and the direction
of the $B$ in the rest frame of the
$D_s^{*}$ candidate. The helicity angle $\theta_V$ is similarly
defined by the direction of the charged daughter particle in the decays
$\rho^+\to\pi^+\pi^0$, $K^{*+}\to K^+\pi^0$, and 
$K^{*+}\to \KS\pi^{+}$. 
Since the momenta of the decay products in
the laboratory frame depend on the helicity angles, acceptance and
efficiency effects modify the ideal angular distribution. We determine
the PDFs of the signal events using 
the Monte Carlo simulation, and measure the angular distribution of
the combinatorial background in the data region $\mes<5.27$~\gevcc.

For \btodspi, \btodsrho, and \btodsk, the fit constrains 
14 free parameters: the shape parameter of the combinatorial background
$\xi$ (1 parameter for all \Ds\ modes), the slopes of the combinatorial
and charmless backgrounds in \mDs\ (3 parameters), the fractions of
true \Ds\ candidates in 
combinatorial background (3), the numbers of combinatorial background
events (3), the numbers of charmless events (3), and the branching fraction
of the signal mode (1). 
In the \btodskstar\ mode (6 individual sub-modes, spanning 3 \Ds\ channels and 
2 $K^{*+}$ channels), 26 free parameters are constrained. 
For the \btodsspi\ and \btodssk\ decays, 5 free
parameters are determined by the fit: $\xi$ (1 parameter for all \Ds\
modes), the number of combinatorial background events (3), and the
branching fraction of the signal mode (1). 
For \btodssrho\ and \btodsskstar\ fits, we add one more free parameter
to the fit: the longitudinal polarization fraction $f_L$ (see
\eqref{eq:helicityshort}). The total number of free parameters is 6 in 
\btodssrho\ and 9 in \btodsskstar. 

\begin{figure*}
\begin{center}
\epsfig{file=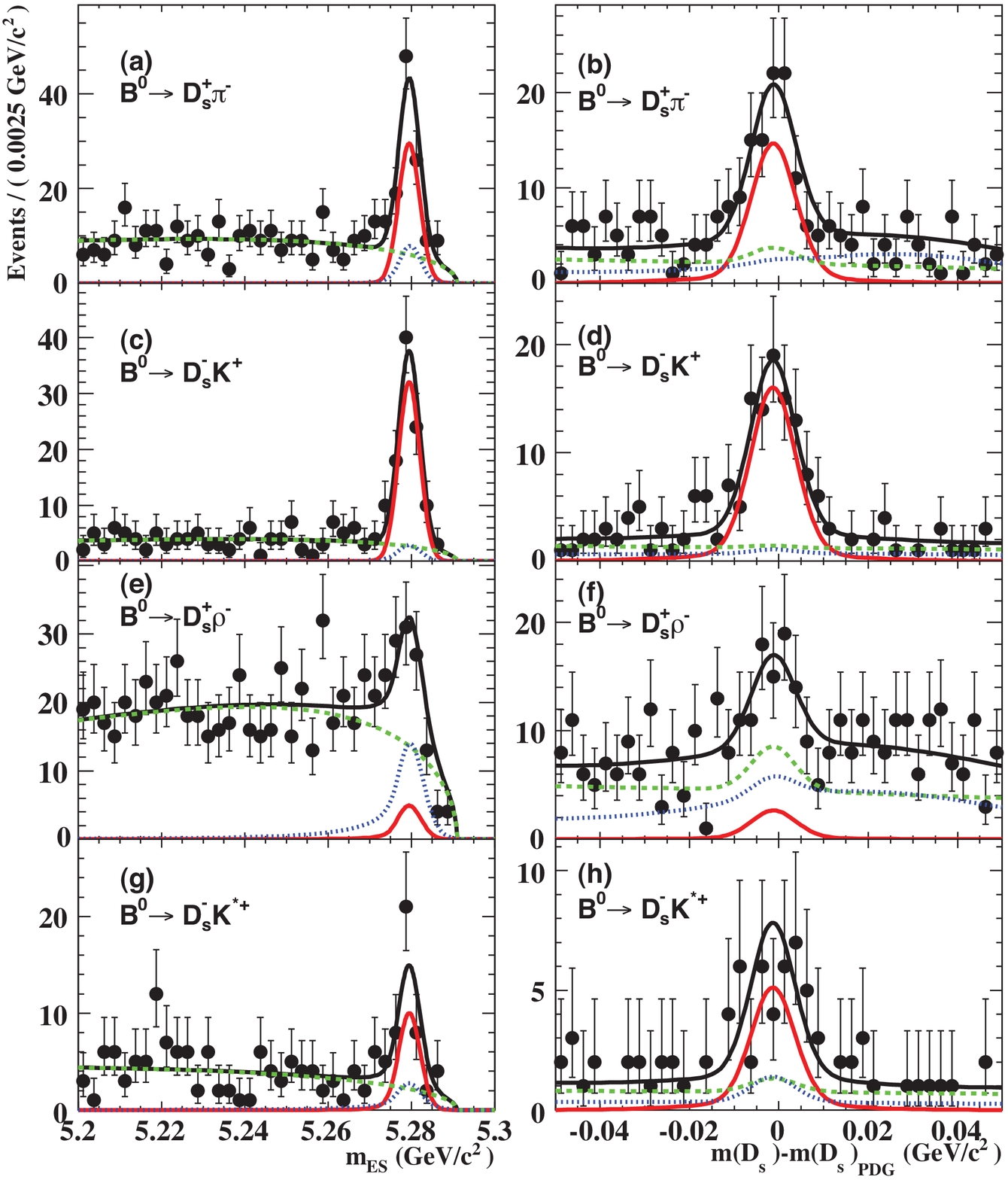,height=7in}
\end{center}
\caption{ (a,c,e,g) \mes\ projection of the fit with 
$|m(\Ds)-m(\Ds)_\mathrm{PDG}|<10$~\mevcc and 
(b,d,f,h) \mDs\ projection with $5.275<\mes<5.285$~\gevcc for (a,b) \btodspi,
(c,d) \btodsk, (e,f) \btodsrho, and (g,h) \btodskstar. The black solid curves
correspond to the full PDF from the combined
fit to all \Ds\ decay modes. Individual contributions are shown 
as solid red (signal), green dashed (combinatorial background),
and blue dotted (sum of reflection, charmless, crossfeed, and
non-resonant backgrounds) curves. 
}
\label{fig:fit1} 
\end{figure*}

\begin{figure*}
\begin{center}
\epsfig{file=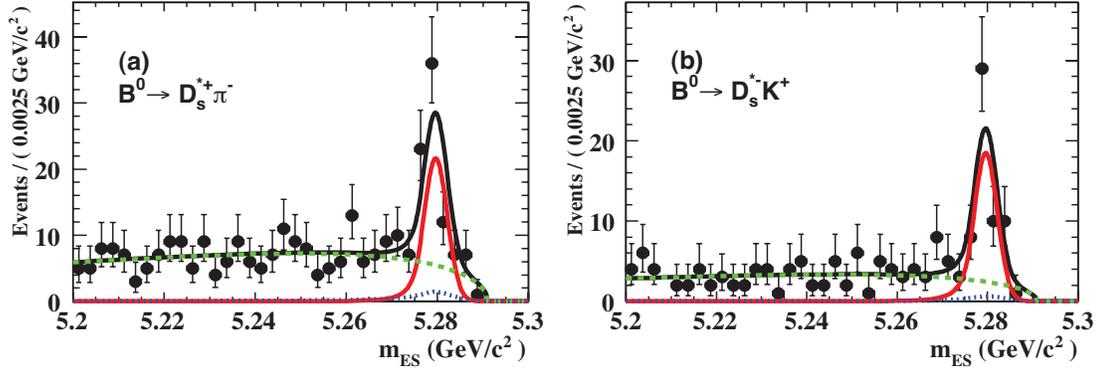,height=2in}
\end{center}
\vspace{0.2cm}
\vspace{-20pt}
\caption{ \mes\ projection of the fit 
for (a) \btodsspi and (b) \btodssk. 
The black solid curves
correspond to the full PDF from the combined
fit to all \Ds\ decay modes. Individual contributions are shown 
as solid red (signal), green dashed (combinatorial background),
and blue dotted (sum of reflection, charmless, and crossfeed
backgrounds) curves. 
}
\vspace{-20pt}
\label{fig:fit2} 
\end{figure*}

\begin{figure*}
\begin{center}
\epsfig{file=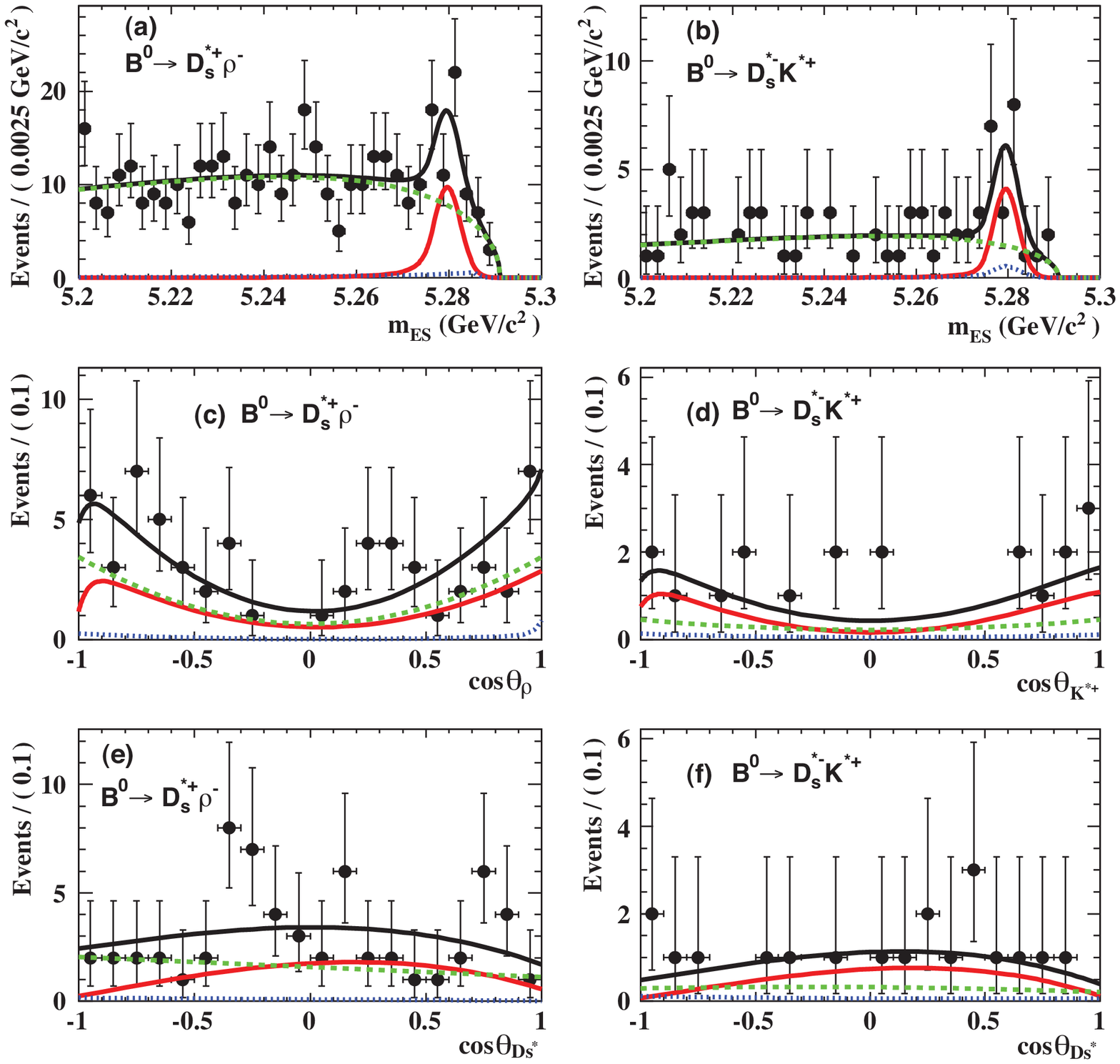,height=6in}
\end{center}
\vspace{-20pt}
\caption{Projections of the fit on to (a,b) \mes, (c) $\cos\theta_\rho$, 
(d) $\cos\theta_{K^*}$, and (e,f) $\cos\theta_{D_s^*}$ for (a,c,e) 
\btodssrho and (b,d,f) \btodsskstar. For helicity projections, a 
selection $5.275<\mes<5.285$~\gevcc is applied. 
The black solid curves
correspond to the full PDF from the combined
fit to all \Ds\ decay modes. Individual contributions are shown 
as solid red (signal), green dashed (combinatorial background),
and blue dotted (sum of reflection, charmless, crossfeed, and non-resonant
backgrounds) curves. 
}
\vspace{-20pt}
\label{fig:fit3} 
\end{figure*}

The results of the fits are shown in
Figs.~\ref{fig:fit1}-\ref{fig:fit3} and summarized in Table~\ref{tab:fit}.

\section{Systematic Uncertainties}

%%% B.F. measurement and systematics

For the branching fractions, the systematic errors are dominated by
the 13\% relative uncertainty for 
$\BR(\Ds\rightarrow\phi\pip)$~\cite{PDG2006}. 
The uncertainty in
the branching fraction ratio
$\BR(\Ds{\to}\Kstarzb\Kp)/\BR(\Ds{\to}\phi\pip)$ 
contributes (2-4)\%, depending on the decay channel. 
For $\BR(\Ds{\to}\KS\Kp)$, we use the most recent 
measurement from the CLEO Collaboration~\cite{ref:KsKCLEO}, which
differs from the previously 
reported central value~\cite{PDG2006} by about 50\%. We
estimate uncertainties due to modeling of the resonance ($K^{*0}$,
$\phi$, $\rho$, and $K^{*+}$) lineshapes by measuring the 
effect of the lineshape variation on signal selection efficiency. 

The uncertainties in the signal selection efficiency are determined by
the accuracy with which the detector effects are modeled in the Monte
Carlo simulations. 
Tracking, particle identification (PID), photon, $\pi^0$ and
$\KS$ reconstruction efficiencies are measured across the wide range
of particle momenta in the dedicated data control
samples. The tracking efficiency and 
resolution are adequately reproduced by the simulations. The simulated
distributions are corrected for the efficiency and resolution of the
$\pi^0$ reconstruction. The efficiency of the $R_L$ cut is also
measured in the data 
control samples, as discussed in Section~\ref{sec:selection}. 

The uncertainties due to the knowledge of the signal and background
PDFs in the ML fit are estimated by measuring the variation of the
fitted values of the branching fractions when PDF parameters are
varied within their uncertainties. 
The correlations between parameters
are taken into account.  
The uncertainties in the signal PDF parameters for the key
discriminants \De, \mes, \mDs, $\Delta m(\Dss)$, and
$\cos\theta_{\Dss}$ are determined by comparing data and Monte
Carlo simulations for the  samples of decays
$\Bz\to D^-\pi^+,\, D^-\rho^+$ ($D^-\to K^+\pi^-\pi^-,\, \KS\pi^-$) 
and $B^+\to\Dbar^{*0}\pi^+,\,\Dbar^{*0}\rho^+$ ($\Dbar^{*0}\to
\Dbar^0\gamma,\, D^0\to K^-\pi^+$). The uncertainties in the signal
PDFs for $\cos\theta_{\rho,K*}$ and the PDFs for the peaking
backgrounds are determined by Monte Carlo simulations. These
distributions depend on 
the modeling of the charged track and $\pi^0$ reconstruction, 
discussed above.  
The helicity angle PDFs for the
continuum background are determined in the data sideband
$m_{ES}<5.27~\gevcc$, and their uncertainties are statistical in
nature. 

Uncertainties due to reflection and crossfeed
backgrounds include the uncertainties in the branching fractions of the
relevant modes, and also account for the contributions of the
sub-dominant modes that are not explicitly included in the ML
fit. These contributions dominate the systematic uncertainty for the
\btodsrho\ mode, which has a small absolute branching fraction. 

As ML estimators may be biased in small samples, 
we measure the bias using a large ensemble of simulated
experiments. In each of these experiments, 
generated according to the sample composition 
observed in data,
the signal and $B$-decay
background events are fully simulated, and the combinatorial
background events are generated from their PDFs. 
The bias is found to be negligible for the 1- and
2-dimensional ML fits (\btodsospi, \btodsosk, \btodskstar\ modes). 
On the other hand, we find that in the vector-vector modes
(\btodssrho\ and \btodsskstar\ decays), the 3-dimensional ML fits
underestimate the true values of the signal branching fraction and
the fraction of the longitudinal polarization. 
We correct for the biases of 
$\Delta\BR=(-0.37\pm 0.03)\times10^{-5}$ and 
$\Delta f_L=(-5.3\pm 0.6)\%$ (\btodssrho) and 
$\Delta\BR=(-0.14\pm 0.04)\times10^{-5}$ and 
$\Delta f_L=(-5.5\pm 0.8)\%$ (\btodsskstar).
We assign a conservative uncertainty of 50\% of
the bias to this correction. 

For the longitudinal polarization fractions $f_L$ in the vector-vector
modes, the systematic errors are dominated by the uncertainties
in the shapes of the signal and background PDFs and the fit bias. 
The systematic uncertainties for each mode are summarized in
Tables~\ref{tab:systematics1}-\ref{tab:systematics3}. 

\section{Results}

We estimate the significance of a non-zero signal yield by computing
$\mathcal{S} = \sqrt{-2\log(\mathcal{L}_0/\mathcal{L}_{\max})}$,
where 
$\mathcal{L}_{\max}$ is the maximum likelihood value, and 
$\mathcal{L}_0$ is the likelihood for a fit in which the signal
contribution is set to zero. 
Including systematic uncertainties and assuming Gaussian-distributed
errors, we obtain signal 
observation significances of 
$3.9$ (\btodssrho), $4.6$ (\btodskstar), and $3.1$ (\btodsskstar) standard
deviations, providing the first evidence for these decays. 
We test that $\mathcal{S}$ measures the probability for the
background events to fluctuate to the observed number of signal events
with a large ensemble of simulated experiments. For each such
experiment, we generate a set of pure background events according to
the PDFs and sample composition observed in our dataset. We then fit
each simulated experiment and measure the signal and background yields
and, for the vector-vector modes, the polarization fraction $f_L$. By
counting the fraction of such pseudo-experiments in which the signal
yields are at least as large as the yield observed in the real dataset,
we confirm that $\mathcal{S}^2$ follows closely the $\chi^2$
distribution with one degree of freedom. 

The branching fraction and polarization results are collected in
Table~\ref{tab:fit}. 
Since we do not
observe a significant event yield in \btodsrho, we set a 90\% 
confidence-level Bayesian upper limit assuming a constant prior for
$\BR(\btodsrho)>0$.  

\section{Conclusions}

We report the following improved measurements of the branching
fractions for the rare 
decays \btodsospi\ and \btodsosk, and the first measurements of the
branching fractions for the decays \btodsosrho\ and \btodsoskstar, as
well as the measurements of
the longitudinal polarization fractions $f_L$ in vector-vector final
states \btodssrho\ and \btodsskstar:
\begin{eqnarray*}
\BR(\btodspi)     &=& [2.5\pm 0.4\pm 0.2]\times 10^{-5}\\
\BR(\btodsspi)    &=& [2.6^{+0.5}_{-0.4}\pm 0.3]\times 10^{-5}\\ 
\BR(\btodsrho)    &=& [1.1^{+0.9}_{-0.8}\pm 0.3]\times 10^{-5}\\
\BR(\btodsrho)    &<& 2.4\times 10^{-5}\ (90\% \mathrm{C.L.})\\
\BR(\btodssrho)   &=& [4.4^{+1.3}_{-1.2}\pm 0.8]\times 10^{-5}\\
f_L(\btodssrho)   &=& 0.86^{+0.26}_{-0.28}\pm 0.15\\
\BR(\btodsk)      &=& [2.9\pm 0.4\pm 0.3]\times 10^{-5}\\
\BR(\btodssk)     &=& [2.4\pm 0.4\pm 0.2]\times 10^{-5}\\
\BR(\btodskstar)  &=& [3.6^{+1.0}_{-0.9}\pm 0.4]\times 10^{-5}\\
\BR(\btodsskstar) &=& [3.0^{+1.4}_{-1.2}\pm 0.3]\times 10^{-5}\\
f_L(\btodsskstar) &=& 0.96^{+0.38}_{-0.31}\pm 0.08\ , \\
\end{eqnarray*}
where the first quoted uncertainty is statistical, and the second is
systematic. 

The branching fractions for \btodsoskokstar\ are small
compared to the dominant decays $\Bz\to D^{(*)-}\pi^+$ and  $\Bz\to
D^{(*)-}\rho^+$, implying 
insignificant contributions from the color-suppressed
$W$-exchange diagrams. The ratios $\BR(\btodsk)/\BR(\btodssk)$ and 
$\BR(\btodskstar)/\BR(\btodsskstar)$ are consistent with the
expectation of unity~\cite{Wexch}. The predictions for the 
branching fractions of \btodsospi\ and \btodsosrho\ decays are based on
the factorization assumption~\cite{ref:factor} and depend on the estimates
of the hadronic form factors. The observed pattern 
$\BR(\btodsrho)<\BR(\btodspi)\approx\BR(\btodsspi)<\BR(\btodssrho)$
appears to be most consistent with the form factors computed in
\cite{ref:FF_BSW}. 
The polarizations of the vector mesons in
\btodssrho\ and \btodsskstar\ are consistent with
expectations~\cite{Wexch,ref:VV}.

Assuming the SU(3) relation, \eqref{eq:rDPi}, and the recent value of 
$f_{D^{(*)}_s}/f_{D^{(*)}}$ from an unquenched Lattice QCD
calculation~\cite{fdsdRef},  
we determine the ratios
of the CKM-suppressed to CKM-favored decay amplitudes in
decays $\B^0\to D^{(*)\pm}\pi^{\mp}$ and $\B^0\to
D^{(*)\pm}\rho^{\mp}$:
\begin{eqnarray*}
r(D\pi)      &=& [1.75\pm0.14\,\stat\pm0.09\,\syst\pm0.10\,(\mathrm{th})]\%\\ 
r(D^{*}\pi)  &=& [1.81^{+0.17}_{-0.14}\,\stat\pm0.12\,\syst\pm0.10\,(\mathrm{th})]\%\\
r(D\rho)     &=& [0.71^{+0.29}_{-0.26}\,\stat\pm0.11\,\syst\pm0.04\,(\mathrm{th})]\%\\
r(D^{*}\rho) &=& [1.50^{+0.22}_{-0.21}\,\stat\pm0.16\,\syst\pm0.08\,(\mathrm{th})]\%\\
\end{eqnarray*}
where the first error is statistical, the second includes experimental
systematics, and the last accounts for the uncertainty in the
theoretical value of $f_{D^{(*)}_s}/f_{D^{(*)}}$~\cite{fdsdRef}. 
These amplitude ratios are below 2\%, 
which implies small $C\! P$ asymmetries in
$\Bz{\to} D^{(*)\mp}\pi^\pm$ and $\Bz{\to} D^{(*)\mp}\rho^\pm$
decays, making it difficult to measure $\sin(2\beta+\gamma)$ precisely
in these
decays. The results presented here 
supersede our previously published measurements~\cite{priorBaBar}. \\

% Standard acknowledgments paragraph; must always be included.
\section{Acknowledgments}

We are grateful for the 
extraordinary contributions of our \pep2\ colleagues in
achieving the excellent luminosity and machine conditions
that have made this work possible.
The success of this project also relies critically on the 
expertise and dedication of the computing organizations that 
support \babar.
The collaborating institutions wish to thank 
SLAC for its support and the kind hospitality extended to them. 
This work is supported by the
US Department of Energy
and National Science Foundation, the
Natural Sciences and Engineering Research Council (Canada),
the Commissariat \`a l'Energie Atomique and
Institut National de Physique Nucl\'eaire et de Physique des Particules
(France), the
Bundesministerium f\"ur Bildung und Forschung and
Deutsche Forschungsgemeinschaft
(Germany), the
Istituto Nazionale di Fisica Nucleare (Italy),
the Foundation for Fundamental Research on Matter (The Netherlands),
the Research Council of Norway, the
Ministry of Education and Science of the Russian Federation, 
Ministerio de Educaci\'on y Ciencia (Spain), and the
Science and Technology Facilities Council (United Kingdom).
Individuals have received support from 
the Marie-Curie IEF program (European Union) and
the A. P. Sloan Foundation.

\onecolumngrid

%% results

%%%%%%%%%%%%%%%%%%%%
%
% TABLES
%
%%%%%%%%%%%%%%%%%%%%
\begin{table}[bh]
\caption{Expected background yields from the dominant $B$ decay modes,
fixed in the likelihood fit.}
\label{tab:bkg}
\begin{center}
\begin{tabular}{ll c c c}
\hline
\hline\\[-0.3cm]
Signal mode & Background mode & $N(\dsphipi)$ & $N(\dskstark)$ & $N(\dsksk)$ \\
\hline\\[-0.3cm]
\btodspi	& \btodsk											& $1.5 \pm 0.2$		& $0.4 \pm 0.1$		& $0.3 \pm 0.0$		\\
			& $B^0\to D^{-}\pi^{+}$								& $17.1 \pm 1.3$	& $21.1 \pm 1.6$	& $13.9 \pm 1.0$	\\
\hline\\[-0.3cm]
\btodsspi	& \btodssk											& $0.9 \pm 0.2$		& $0.3 \pm 0.1$		& $0.3 \pm 0.0$	\\
			& $B^0\to D^{-}\rho^{+}$							& $0.5 \pm 0.1$		& $3.5 \pm 0.7$		& $1.8 \pm 0.4$	\\
			& $B^0\to D^{*-}\pi^{+}, D^{*-}\to D^-\pi^0$		& $0.3 \pm 0.1$		& $1.2 \pm 0.2$		& $0.8 \pm 0.1$	\\
\hline\\[-0.3cm]
\btodsrho	& \btodssrho										& $6.9 \pm 2.0$		& $1.4 \pm 0.4$		& $1.6 \pm 0.4$	\\
			& \btodsspi											& $6.3 \pm 1.1$		& $1.3 \pm 0.2$		& $1.6 \pm 0.3$	\\
			& $B^+\to D_s^{*+}\pi^0$							& $3.9 \pm 0.7$		& $1.1 \pm 0.2$		& $1.0 \pm 0.2$	\\
			& $B^0\to D^{-}\rho^{+}$							& $26.0 \pm 4.4$	& $35.2 \pm 5.9$	& $30.1 \pm 5.0$ \\
			& $B^0\to D^{*-}\rho^{+}, D^{*-}\to D^0\pi$			& $0.3 \pm 0.0$		& $6.1 \pm 3.7$		& $8.5 \pm 1.4$	\\
			& $B^0\to D^{*-}\rho^{+}, D^{*-}\to D^-\pi^0$		& $0.9 \pm 0.4$		& $1.5 \pm 0.6$		& $2.2 \pm 0.5$	\\
\hline\\[-0.3cm]
\btodssrho	& $B^0\to D^{-}\rho^{+}$							& $0.7 \pm 0.2$		& $1.7 \pm 0.4$		& $2.6 \pm 1.2$	\\
			& $B^0\to D^{*-}\rho^{*+}, D^{*-}\to D^{-}\pi^0$	& $0.1 \pm 0.0$		& $0.8 \pm 0.1$		& $0.8 \pm 0.1$	\\
\hline\\[-0.3cm]
\btodsk		& \btodspi											& $0.6 \pm 0.1$		& $0.3 \pm 0.0$		& $0.2 \pm 0.0$		\\
			& \btodssk											& $1.4 \pm 0.3$		& $0.3 \pm 0.1$		& $0.2 \pm 0.0$		\\
			& $B^0\to D^{-}K^{+}$								& $0.9 \pm 0.3$		& $2.2 \pm 0.2$		& $1.3 \pm 0.4$		\\
\hline\\[-0.3cm]
\btodssk	& \btodsk											& $0.9 \pm 0.1$		& $0.2 \pm 0.0$		& $0.1 \pm 0.0$	\\
			& $B^0\to D_s^{-}K^{*+}$							& $1.0 \pm 0.2$		& $0.4 \pm 0.1$		& $0.3 \pm 0.1$	\\
			& \btodsspi											& $0.5 \pm 0.1$		& $0.2 \pm 0.0$		& $0.2 \pm 0.0$	\\
\hline\\[-0.3cm]
\btodskstar, $K^{*+}\to \KS\pi^+$
			& $B^0\to D_s^{*-}K^{*+}$							& $0.4 \pm 0.2$		& $0.1 \pm 0.0$		& $0.1 \pm 0.0$	\\
			& $B^0\to D_s^{-}\pi^+K^0$							& $1.9 \pm 0.7$		& $0.8 \pm 0.1$		& $0.6 \pm 0.1$	\\
\btodskstar, $K^{*+}\to K^+\pi^0$ 
			& $B^0\to D_s^{*-}K^+$								& $2.6 \pm 0.4$		& $0.9 \pm 0.2$		& $0.9 \pm 0.2$	\\
			& $B^0\to D_s^{*-}K^{*+}$							& $1.1 \pm 0.5$		& $0.3 \pm 0.2$		& $0.4 \pm 0.2$	\\
			& $B^0\to D_s^{-}K^+\pi^0$							& $0.4 \pm 0.4$		& $0.1 \pm 0.1$		& $0.1 \pm 0.1$	\\
\hline\\[-0.3cm]
\btodsskstar,  $K^{*+}\to \KS\pi^+$ 
			& $B^0\to D_s^{-}K^{*+}$							& $0.2 \pm 0.0$		& $0.1 \pm 0.0$		& $0.0 \pm 0.0$	\\
			& $B^0\to D_s^{*-}\pi^+K^0$							& $0.6 \pm 0.4$		& $0.3 \pm 0.2$		& $0.2 \pm 0.1$	\\
\btodsskstar, $K^{*+}\to K^+\pi^0$ 
			& $B^0\to D_s^{-}K^{*+}$							& $0.5 \pm 0.1$		& $0.1 \pm 0.0$		& $0.2 \pm 0.1$	\\
			&  $B^0\to D_s^{*-}K^+\pi^0$						& $0.1 \pm 0.1$		& $0.0 \pm 0.0$		& $0.0 \pm 0.0$	\\
\hline
\hline
\end{tabular}
\end{center}
\end{table}

\begin{table}[bh]
\caption{The number of reconstructed candidates ($N_\mathrm{raw}$), 
the  signal yield ($N_\mathrm{sig}$), computed from the fitted
branching fractions, the combinatorial background
($N_\mathrm{comb}$), and the sum of charmless, reflection, non-resonant, and
crossfeed contributions ($N_\mathrm{peak}$), 
extracted from the likelihood fit. 
Also given are 
the reconstruction efficiency ($\varepsilon$), 
 the signal significance $\mathcal{S}$,  
the measured branching fraction \BR, and the fraction of 
longitudinal polarization $f_L$ (where appropriate). 
The first uncertainty is  statistical, and the second is systematic. 
}

\begin{center}
\begin{tabular*}{\textwidth}{@{\extracolsep{\fill}} ll c c c c c c c c} \hline \hline
$B$ mode& $D_s$ mode & $N_\mathrm{raw}$ & $N_\mathrm{sig}$  & $N_\mathrm{comb}$ & $N_\mathrm{peak}$ & 
$\varepsilon$(\%)& $\mathcal{S}$ & \BR ($10^{-5}$) & $f_L$ \\
\hline
           & $\Ds{\to}\phi\pip$    & $582$ &$51\pm10$ &$500\pm24$ &$32\pm10$ & $25.2$ & & & \\
$\btodspi$ & $\Ds{\to}\Kstarzb\Kp$ & $402$ &$19\pm4$ &$352\pm20$ &$36\pm8$ &$8.0$ &$8.2\sigma$ & $2.5\pm0.4\pm0.2$ &  \\
           & $\Ds{\to}\KS\Kp$      & $282$ &$19\pm4$ &$245\pm16$ &$25\pm7$ &$19.4$ & & & \\
\hline
           & $\Ds{\to}\phi\pip$    & $150$ &$34\pm6$ &$113\pm12$ & $1.7\pm0.3$ &$16.7$ & & & \\
$\btodsspi$&$\Ds{\to}\Kstarzb\Kp$  & $96$ &$13\pm2$ &$77\pm9$ & $5.0\pm0.7$ &$5.5$ & $6.8\sigma$& $2.6^{+0.5}_{-0.4}\pm0.3$ & \\
           & $\Ds{\to}\KS\Kp$      & $52$ &$13\pm2$ &$41\pm7$ & $2.9\pm0.4$ &$13.2$ & & & \\
\hline
           & $\Ds{\to}\phi\pip$    & $1190$ &$11\pm9$ &$1102\pm36$ &$78\pm17$  &$12.1$ & & & \\
$\btodsrho$ & $\Ds{\to}\Kstarzb\Kp$ & $644$ &$3\pm3$ &$584\pm26$  &$59\pm13$ &$3.3$ &$1.3\sigma$ & $1.1^{+0.9}_{-0.8}\pm0.3$ &  \\
           & $\Ds{\to}\KS\Kp$      & $613$ &$4\pm4$ &$544\pm25$ &$70\pm13$ &$8.3$ & & $<2.4$ (90\% C.L.)& \\
\hline
			&$\Ds{\to}\phi\pip$    & $194$	&$22\pm6$ 	&$175\pm14$	& $0.8\pm0.2$	&$6.3$ & & & \\
$\btodssrho$&$\Ds{\to}\Kstarzb\Kp$ & $101$	&$7\pm2$ 	&$93\pm10$	& $2.5\pm0.4$	&$1.9$ & $3.9\sigma$& $4.4^{+1.3}_{-1.2}\pm0.8$ & $0.86^{+0.26}_{-0.28}\pm 0.15$ \\
			&$\Ds{\to}\KS\Kp$      & $91$	&$8\pm2$ 	&$80\pm10$	& $3.4\pm1.2$	&$4.6$ & & & \\
\hline
          & $\Ds{\to}\phi\pip$    & $307$ &$55\pm8$ &$240\pm16$ &$15\pm7$  &$22.9$ & & & \\
$\btodsk$ & $\Ds{\to}\Kstarzb\Kp$ & $262$ &$23\pm3$ &$227\pm16$ &$11\pm6$  &$8.2$ & $11\sigma$& $2.9\pm0.4\pm0.3$ &  \\
          & $\Ds{\to}\KS\Kp$      & $148$ &$20\pm3$ &$125\pm12$ &$6\pm4$   &$17.4$ & & & \\
\hline
           & $\Ds{\to}\phi\pip$    & $76$ &$28\pm5$ &$47\pm8$ & $2.4\pm0.3$ & $15.2$  &&&\\
$\btodssk$ & $\Ds{\to}\Kstarzb\Kp$ & $50$ &$12\pm2$ &$39\pm7$ & $0.8\pm0.1$ & $5.7$ &$7.4\sigma$& $2.4\pm0.4\pm0.2$ &  \\
           & $\Ds{\to}\KS\Kp$      & $34$ &$14\pm2$ &$21\pm5$ & $0.6\pm0.1$ & $12.0$  &&&\\
\hline
$\btodskstar$ & $\Ds{\to}\phi\pip$				& $95$ &$9\pm3$ &$83\pm10$ &$4\pm4$  &$13.8$ & & & \\
$K^{*+}\to \KS\pi^+$ & $\Ds{\to}\Kstarzb\Kp$	& $45$ &$4\pm1$ &$40\pm7$  &$1\pm2$  &$5.4$  & & &  \\
          & $\Ds{\to}\KS\Kp$					& $33$ &$3\pm1$ &$27\pm6$  &$1\pm3$  &$10.2$ & & & \\
        & $\Ds{\to}\phi\pip$					& $157$ &$9\pm3$ &$150\pm13$ &$1\pm4$   &$9.0$ & $4.6\sigma$ & $3.6^{+1.0}_{-0.9}\pm0.4$ & \\
$K^{*+}\to K^+\pi^0$ & $\Ds{\to}\Kstarzb\Kp$	& $94$  &$3\pm1$ &$83\pm10$  &$6\pm4$   &$3.1$ & & &  \\
          & $\Ds{\to}\KS\Kp$					& $96$  &$3\pm1$ &$83\pm10$  &$9\pm4$  &$7.1$ & & & \\
\hline
$\btodsskstar$ & $\Ds{\to}\phi\pip$					& $16$ &$4\pm 2$ &$14\pm4$ & $0.8\pm 0.4$ & $6.7$  &&&\\
$K^{*+}\to \KS\pi^+$ & $\Ds{\to}\Kstarzb\Kp$		& $8$  &$2\pm 1$ & $7\pm3$ & $0.4\pm 0.2$ & $2.5$  &&&\\
           & $\Ds{\to}\KS\Kp$						& $7$  &$1\pm 1$ & $6\pm3$ & $0.2\pm 0.1$ & $5.2$  &&&\\
$\btodsskstar$ & $\Ds{\to}\phi\pip$					& $30$ &$4\pm 2$ &$22\pm6$ & $0.6\pm 0.2$ & $5.1$  & $3.1\sigma$ & $3.0^{+1.4}_{-1.2}\pm 0.3$& $0.96^{+0.38}_{-0.31}\pm 0.08$\\
$K^{*+}\to K^+\pi^0$ & $\Ds{\to}\Kstarzb\Kp$		& $3$  &$2\pm 1$ & $3\pm2$ & $0.1\pm 0.0$ & $1.8$  &&&\\
           & $\Ds{\to}\KS\Kp$						& $11$ &$2\pm 1$ & $9\pm3$ & $0.2\pm 0.1$ & $4.1$  &&&\\
\hline\hline
\end{tabular*}
\end{center}
\label{tab:fit}
\end{table}

%
% SYSTEMATICS 
%
%
\begin{table}[ht]
\begin{center}
\caption{Relative systematic uncertainties for the branching fractions of $B^0\to D_sX$ modes (\%).}
\label{tab:systematics1}
\begin{tabular}{c c c c c}
\hline
\hline
						& \btodspi & \btodsk & \btodsrho & \btodskstar \\
\hline
$N_{\BB}$					& 1.1 & 1.1 & 1.1 & 1.1 \\
Tracking efficiency			& 1.6	& 1.3	& 1.7	& 2.0		\\
PID efficiency				& 1.0 & 1.0 & 1.0 & 1.0 \\
$\pi^0$ efficiency			& -		& -		& 3.0	& 3.0		\\
$R_L$ cut efficiency		& 1.6	& 1.4	& 2.8	& 2.0		\\
MC statistics				& 0.8	& 0.7	& 1.8	& 1.1		\\
\KS\ efficiency				& 0.4	& 0.3	& 0.9	& 0.0		\\
PDF parameters  			& 0.8	& 0.7	& 2.8	& 1.4		\\
$\De$, $m({D_s})$ PDFs		& 0.6	& 1.7	& 3.2	& 1.7		\\
\BR(\dsphipi)				& 7.7	& 8.3	&14.7	& 7.7		\\
\BR(\dskstark)				& 2.8	& 3.1	& 1.8	& 3.1		\\
\BR(\dsksk)					& 1.6	& 1.4	& 3.7	& 0.6		\\
Reflection background		& 2.0	& 0.7	&10.2	& 0.6		\\
Crossfeed background		& 0.9	& 0.4	&15.6	& 2.5		\\
Resonant lineshape			& 1.6	& 1.7	& 1.8	& 4.9		\\
\hline
{\bf TOTAL}					& 9.3	& 9.7	& 25.0	& 13.1		\\
\hline
\hline
\end{tabular} 
\end{center}
\end{table}
\begin{table}[hbt]
\begin{center}
\caption{Relative systematic uncertainties for the branching fractions
  of $B^0\to D_s^*X$ modes (\%). }
\label{tab:systematics2}
\begin{tabular}{c c c c c}
\hline
\hline
							& \btodsspi & \btodssk & \btodssrho & \btodsskstar\\
\hline
$N_{\BB}$				& 1.1 & 1.1 & 1.1 & 1.1 \\
Tracking efficiency			& 1.6	& 1.3	&	1.7		& 2.0	\\
PID efficiency				& 1.0 & 1.0 & 1.0 & 1.0 \\
$\pi^0$ efficiency			& -		& -		& 3.0		& 3.0	\\
$\gamma$ efficiency			& 1.8 & 1.8 & 1.8 & 1.8 \\
\BR(\dssdsgam)				& 0.7 & 0.7 & 0.7 & 0.7 \\
$\Delta m(\Dss)$			& 1.0 & 1.0 & 1.0 & 1.0 \\
$\cos\theta_{\Dss}$			& 1.0 & 1.0 & 1.0 & 1.0 \\
$\pi^0$ veto efficiency		& -		& -		& 2.0		& 2.0	\\
Fit bias					& -		& -		& 4.1		& 2.3	\\
$R_L$ cut efficiency		& 1.5	& 1.7	& 2.3		& 1.6	\\
MC statistics				& 0.8	& 0.8	& 1.1		& 1.6	\\
\KS\ efficiency				& 0.4	& 0.0	& 0.2		& 0.3	\\
PDF parameters				& 1.9	& 2.1	& 3.0		& 3.1	\\
$\De$, $m({D_s})$ PDFs		& 3.0	& 0.8	& 0.7		& 3.4	\\
\BR(\dsphipi)				& 7.6	& 8.4	& 16.0		& 5.9	\\
\BR(\dskstark)				& 2.7	& 3.8	& 3.4		& 4.1	\\
\BR(\dsksk)					& 1.5	& 1.3	& 0.9		& 1.3	\\
Reflection background		& 3.4	& 0.0	& 2.1		& 0.0	\\
Crossfeed background		& 1.9	& 0.4	& 0.0		& 1.6	\\
Resonant lineshape			& 0.8	& 0.4	& 1.6		& 0.3	\\
\hline
{\bf TOTAL}					& 10.5	& 10.2	&18.2		& 10.6	\\
\hline
\hline
\end{tabular} 
\end{center}
\end{table}
\begin{table}[hbt]
\begin{center}
\caption{Absolute systematic uncertainties for the longitudinal polarization fraction (\%).}
\label{tab:systematics3}
\begin{tabular}{c c c}
\hline
\hline
							& \btodssrho & \btodsskstar \\
\hline
Fit bias					& 2.7 & 2.8 \\
$R_L$ cut efficiency		& 0.7 & 0.2 \\
MC statistics				& 0.6 & 0.7 \\
\KS\ efficiency				& 0.4 & 0.0 \\
PDF parameters	            & 14.7& 7.0 \\
$\De$, $m({D_s})$ PDFs	    & 0.0 & 0.5 \\
\BR(\dsphipi)				& 0.8 & 0.8 \\
\BR(\dskstark)				& 0.7 & 0.2 \\
\BR(\dsksk)					& 0.5 & 0.2 \\
Reflection background		& 0.5 & 0.0 \\
Crossfeed background	    & 0.0 & 0.6 \\
Resonant lineshape			& 0.6 & 0.1	\\
\hline
{\bf TOTAL}					&15.0 & 7.6 \\
\hline
\hline
\end{tabular} 
\end{center}
\end{table}

%%%%%%%%%%%%%%%%%%%%
%
% FIGURES
%
%%%%%%%%%%%%%%%%%%%%

%


\begin{thebibliography}{99}
\bibitem{chconj}
Charge conjugation is implied throughout this letter, unless
explicitly stated.
\end{thebibliography}
\end{document}